\documentclass[%
 reprint,
superscriptaddress,
 amsmath,amssymb,
 aps,
prm,
]{revtex4-1}

\usepackage{graphicx}
\usepackage{epstopdf}
\usepackage{dcolumn}
\usepackage{bm}
\usepackage{gensymb}
\usepackage{upgreek}
\usepackage{hyperref}
\usepackage{tabularx}

\preprint{APS/123-QED}

\begin{document}

\title{A new room-temperature equation of state of Bi up to 260 GPa}

\author{Daniel J. Campbell}
\author{Daniel T. Sneed}
\author{E.F. O'Bannon III}
\author{Per S\"{o}derlind}
\author{Zsolt Jenei}
\affiliation{Lawrence Livermore National Laboratory, 7000 East Avenue, Livermore, CA 94550, USA}



\date{\today}

\begin{abstract}

At room temperature, bismuth undergoes several structural transitions with increasing pressure before taking on a body-centered cubic (bcc) phase at approximately 8~GPa. The bcc structure is stable to the highest measured pressure and its simplicity, along with its high compressibility and atomic number, make it an enticing choice as a pressure calibrant. We present three data sets on the compression of bismuth in a diamond anvil cell in a neon pressure medium, up to a maximum pressure of about 260~GPa. The use of a soft pressure medium reduces deviatoric stress when compared to previous work. With an expanded pressure range, higher point density, and a decreased uniaxial stress component, we are able to provide more reliable equation of state (EOS) parameters. We also conduct density functional theory (DFT) electronic-structure calculations that confirm the stability of the bcc phase at high pressure.

\end{abstract}

\maketitle

\section{Introduction}

Elemental bismuth has been the subject of extensive investigation at high pressure. At ambient temperature, it has a series of structural transitions under 10~GPa, passing through four different phases, including a complex host-guest phase~\cite{McMahonHostGuest,DegtyarevaHighPGroup15,HusbandBidDAC}. The final structure, body-centered cubic, emerges at around 8~GPa and has been shown to be stable through at least 200~GPa~\cite{AkahamaBiEOS}. Dynamic compression work has shown that these phase boundaries are sensitive to compression rate. The II and III phases may not appear in shock compression experiments~\cite{HuBiShock,GormanBiSolidSolid,PepinBiKinetics} and other solid-solid or solid-liquid boundaries can move as much as several GPa depending on compression rate~\cite{GormanBiMelting,HusbandBidDAC}. 

Despite the complexity of the lower presure phase diagram, the The stability of bcc Bi-V to high pressure is unquestioned. This simple structure is one reason why Bi has been an enticing choice for a pressure calibrant into the Mbar range. Bi also has the highest atomic number of any non-radioactive element, resulting in a strong x-ray diffraction (XRD) signal. In addition to that, it is softer than many of the other materials (Au, Cu, Pt, NaCl) whose lattice parameters are used for determining pressure in XRD experiments. This means that the volume change with pressure is larger. This increases the precision of pressure determination, which is especially important at ultrahigh compression where the pressure-volume curve flattens out.

Despite the many reports on the structural transitions of Bi, only a limited number have extended far into the bcc phase. The highest pressure work, which reached 220~GPa, did not used a pressure transmitting medium (PTM), counting on Bi itself to redistribute anisotropic stress~\cite{AkahamaBiEOS}. We have carried out x-ray diffraction experiments on bismuth at ambient temperature up to 260 GPa, using neon as a soft pressure medium. With both Cu and Ne as reference pressure calibrants, we find good agreement between three independent data sets. Differences in measured Bi volume at high pressure from previous reports are evidence that a soft pressure medium is necessary to minimize the uniaxial stress on the sample. In this way we are able to provide a more accurate Bi equation of state (EOS), extended to higher pressure. We also show the importance of a soft PTM in deriving accurate EOS values. The data presented in this manuscript will enable Bi to be used for precise pressure calibration at high pressures.

\section{Methods}

\begin{figure*}
    \centering
    \includegraphics[width=0.8\textwidth]{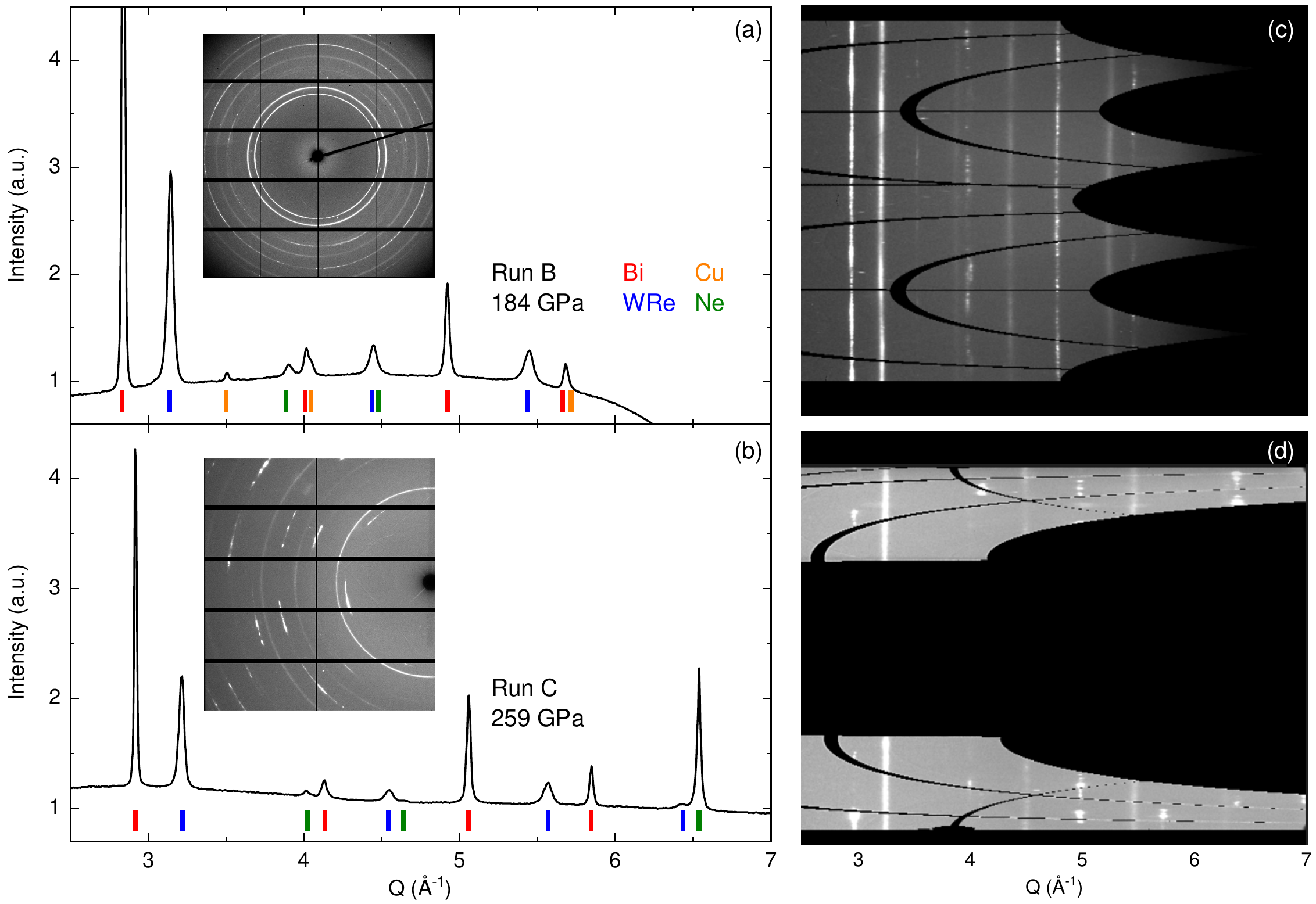}
    \caption{The intensity as a function of $Q$ at (a) 184 and (b) 259~GPa for runs B and C, respectively. These represent the highest pressures reached in each experiment. Ticks below the data mark the peaks for each material present (note that B had a Cu pressure marker, and C did not). The insets show the original two dimensional detector images. (c) and (d) are the corresponding cake images, created by unrolling the detector image to present it as a function of \textit{Q} with the azimuthal angle along the \textit{y}-axis.}
    \label{fig:XRD}
\end{figure*}

Experiments were carried out at Sector 16 of the Advanced Photon Source, as part of the High-Pressure Collaborative Access Team (HPCAT), on three different samples loaded in diamond anvil cells. A set each of 200~$\upmu{}$m flat and 100~$\upmu{}$m/300~$\upmu{}$m beveled culet anvils were used for experiments at station 16-BMD with 25~keV x-rays, and 50~$\upmu{}$m/300~$\upmu{}$m beveled culets at 16-IDB with 30~keV x-rays. We will refer to the different experiments as runs A, B, and C, in decreasing order of culet size. All samples were loaded with bismuth foil (Goodfellow, 99.97\%, 5~$\upmu$m thickness) and neon gas via a high pressure gas loader.  The 200~$\upmu{}$m culet DAC also contained copper foil (Goodfellow, 99.97\%, 5~$\upmu$m thickness) and ruby for pressure calibration, while the 100~$\upmu{}$m DAC contained a copper sphere approximately 10~$\upmu{}$m in diameter. Gaskets were indented to thicknesses of about 25~$\upmu{}$m (A and B) and 20~$\upmu{}$m (C), and holes were drilled in the center of the gaskets with diameters of 120, 55, and 20~$\upmu{}$m for A, B, and C, respectively. The maximum pressure reached in each experiment was, in order 89, 184, and 259~GPa. Pressure was increased during experiments via a helium gas membrane. High quality x-ray diffraction patterns, with good signal to noise ratio, were collected up to the highest pressures, as demonstrated in Fig.~\ref{fig:XRD}, which shows the 2D detector image, caked diffraction image, and intensity as a function of $Q$ for the highest pressure points of Runs B and C. Note that the detector configuration was the same in Runs A and B but different in Run C, where it covered less \textit{Q} than in the other measurements.

Several crystal structures were considered in our theoretical modeling of Bi, namely Bi-I (rhombohedral), Bi-II (monoclinic), simple cubic (sc), hexagonal close-packed (hcp) with either an ideal $c/a$ axial ratio (1.633) or an optimized (relaxed) axial ratio, face-centered cubic (fcc), and finally the Bi-V structure (bcc). Because we are primarily focused in this report on the high-pressure behavior, we do not consider the low-pressure Bi-III phase, which has a complex host-guest structure~\cite{McMahonHostGuest}. Also, full structural relaxations are not performed for Bi-I and Bi-II\textemdash{}we instead use the experimentally reported structural parameters. We apply DFT with the generalized gradient approximation (GGA) for the electron exchange and correlation functional. The methodology is implemented in an all-electron code to avoid the commonly used pseudopotential approximation that is often assumed in DFT calculations but tends to cause inaccuracies at very high compression. Specifically, we employ a full-potential linear muffin-tin orbital (FPLMTO) method~\cite{fplmto} that has been shown to be very accurate for high-pressure studies for many materials including bismuth up to 1 Mbar~\cite{SoderlindYoung}.

The FPLMTO technique does not make any approximations beyond that of the GGA and limitations of the basis set. Basis functions, electron densities and potentials are calculated without any geometrical approximation and these are expanded in spherical harmonics inside non-overlapping (muffin-tin) spheres surrounding each atom and in Fourier series in the region between these muffin-tin spheres. There is a choice in how to define the muffin-tin sphere radius and here it is chosen as 0.74 of the radius of a sphere with a volume equal the atomic volume (Wigner-Seitz radius). The radial part of the basis functions inside the muffin-tin spheres are calculated from a wave equation for the $l = 0$ component of the potential that include all relativistic corrections including spin-orbit coupling for d and f states but not for the p states, following the comprehensive discussions of the spin-orbit interaction in~\cite{nordstrom,sadigh}.
All calculations utilize semi-core states and valence states with two fixed energy parameters each for the s semi-core state, p semi-core state and the valence states. There are six tail energies ranging from -3 to -0.2 Ry. Furthermore, we define 14 basis functions with 5s, 5p, and 4f semi-core states in addition to the 6s, 6p, 5d, and 5f valence states.
The number of k-points included in the electronic-structure calculations depend on the particular crystal structure but we generally use about 1000 k-points or more for one atom/cell calculations and less for cells with many atoms. Each energy eigenvalue is broadened with a Gaussian having a width of 20 mRy. Total energies, for each phase, were calculated on a dense volume mesh of about 0.3 ~\AA$^3$.

\section{Pressure-Volume Relation}

\begin{figure}
    \centering
    \includegraphics[width=0.48\textwidth]{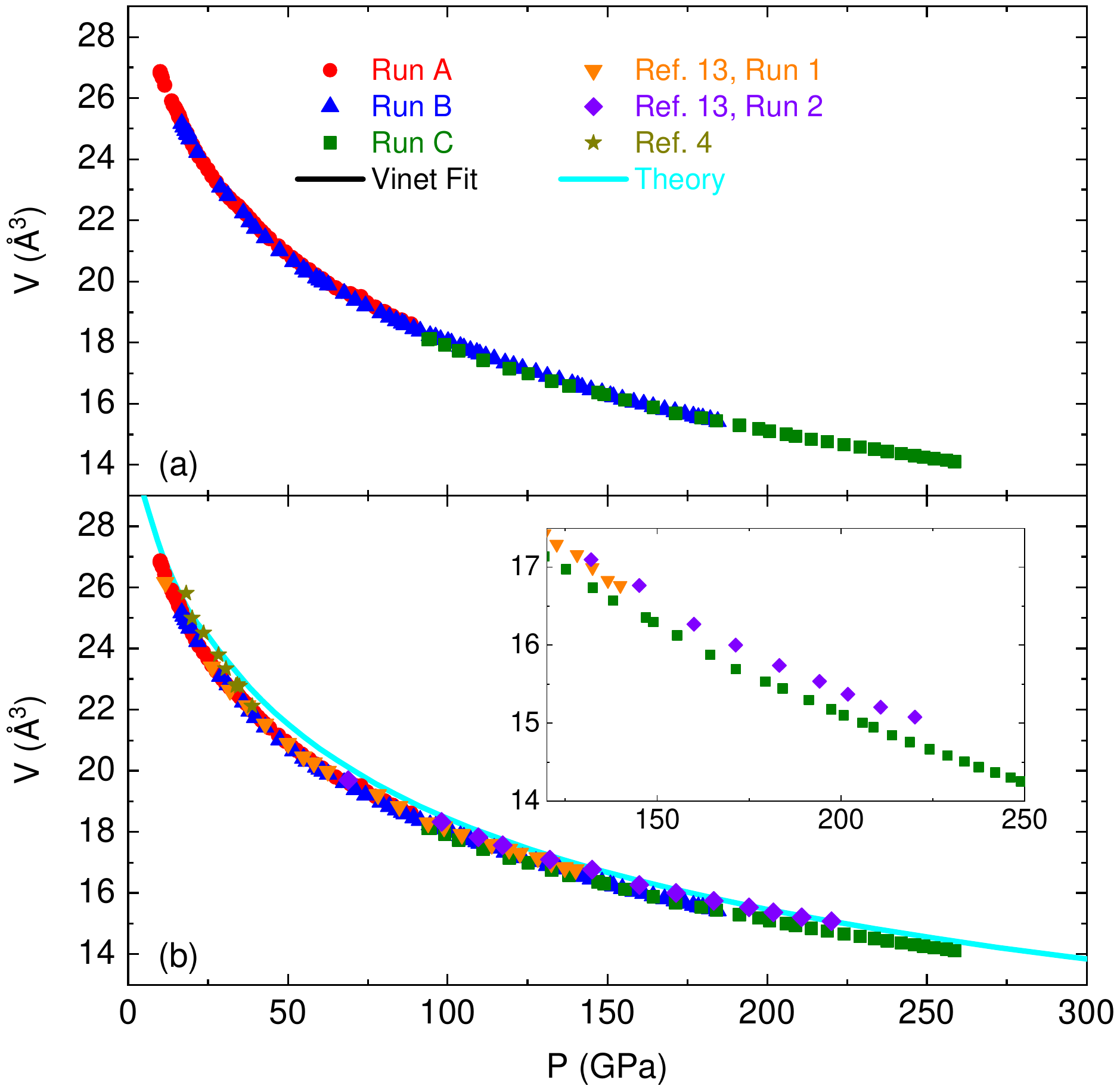}
    \caption{(a) Pressure-volume relationship for Bi-V obtained from our three experiments, with 200 (red circles), 100/300 (blue triangles), and 50/300 (green squares)~$\upmu{}$m culets. The solid line is a fit to all three data sets using the Vinet equation of state with $K_0 = 37.7$~GPa, $K'_0 = 5.9$, and $V_0 = 31.67$~\AA{}$^3$. (b) The same data in (a) compared to data from Refs.~\cite{AkahamaBiEOS,LiuBiEOS} and our own theoretical calculations (solid line). The lower volume for a given pressure in our data is what would be expected for reduced deviatoric stress. The inset shows the highest pressure region for Run C and the two runs of Ref.~\cite{AkahamaBiEOS}.}

\label{fig:PVAll}
\end{figure}

At ambient temperature, Bi-I transitions to Bi-II, Bi-III, and finally Bi-V, while Bi-IV appears only above room temperature~\cite{DegtyarevaHighPGroup15}. We saw evidence for all four ambient temperature structures during initial pressure increase, with some coexistence of Bi-III and Bi-V past the latter's first appearance at around 8~GPa. Above 10 GPa in all runs, only Bi-V was observed in the pattern, and so all Bi-V data used for analysis start at 10~GPa. For runs A and B the position of the Cu [111] reflection was used to calculate pressure~\cite{DewaeleEOSSixMetals}  For Run C pressure was determined using the Ne [111] peak. Our Ne EOS was calculated using the results of the 100~$\upmu{}$m culet experiment, by calibrating the lattice parameter as determined by the Ne [111] reflection to the pressure as determined by Cu. Using a Vinet EOS form,

\begin{equation}
    P = 3K_0\Big(\frac{1-(V/V_0)^{\frac{1}{3}}}{(V/V_0)^{\frac{2}{3}}}\Big)e^{\frac{3}{2}(K_0'-1)(1-(V/V_0)^{\frac{1}{3}})}
    \label{VinetEOS}
\end{equation}

and fixing the atomic volume $V_0$ = 22.234~\AA$^3$ as in Ref.~\cite{DewaeleNeDiamondEOS} we obtain a bulk modulus $K_0$ = 1.05~GPa and its pressure derivative $K_0'$ = 8.38. The process of calibrating Ne to our own Cu data minimizes the uncertainty that could be introduced by using multiple pressure markers with separately determined equations of state. This method ensures that the pressure values in our data derive from the same source with the fewest intermediate steps possible. While they are independent of each other, preparation for Runs B and C was very similar. Our calibration also extends to 184~GPa, while some other commonly used Ne equations of state have only been determined up to around 100~GPa~\cite{FeiNeEOS}. Our EOS parameters are very similar to the parameters determined by Dewaele et al. up to 209~GPa, ($K_0 = 1.070$~GPa, $K_0'= 8.40$, with $V_0$ set to be equal)~\cite{DewaeleNeDiamondEOS} [Supplemental Material, Table~S-I]. However, as shown in the Supplemental Material [Fig.~S1] even the slight difference between the two leads to a 10~GPa difference at the maximum pressure, and a clear difference in curvature when comparing multiple data sets.

Figure~\ref{fig:PVAll}(a) shows the volume of Bi as a function of pressure for all three of our data sets, starting at 10~GPa when we observe only the bcc phase. The extracted lattice parameters for Bi-V, Cu, and Ne for all runs are available in the SM [Table~S-II]. Maximum pressures reached were 89, 184, and 259~GPa for the 200, 100/300, and 50/300~$\upmu{}$m culet cells, respectively. Note that for Run C, few data points were obtained below 100~GPa. Figure~\ref{fig:PVAll}(b) shows a comparison to the previous data~\cite{AkahamaBiEOS,LiuBiEOS}. For the experiments of Ref.~\cite{AkahamaBiEOS} no PTM was used, with only Bi foil and Pt (the pressure calibrant) in the cell. We would expect this to result in more anisotropic stress than in experiments with Ne as the PTM. While Ne solidifies above 4~GPa at ambient pressure, it is still much more compressible than Bi. Indeed, the upward deviation of the data of Ref.~\cite{AkahamaBiEOS} at high pressure, clearest in the inset to Fig.~\ref{fig:PVAll}(b), is what would be expected under less hydrostatic conditions. Our three data sets also combine to give a much higher point density over the covered range.

We fit an EOS of the form given in Eq.~\ref{VinetEOS} to extract $K_0$, $K_0'$, and $V_0$. One issue with Bi is that, since the bcc structure is not stable at ambient pressure, the $V_0$ of Bi-V is not known, and small variations dramatically change the corresponding $K_0$ and $K_0'$. We calculated the bulk modulus and its derivative for two different $V_0$ values: 32.23 and 31.67~\AA$^3$. 32.23~\AA$^3$ is the theoretical ambient pressure volume of bcc Bi. The value of 31.67~\AA$^3$ is the result of letting all parameters of the fit vary. Notably, this value is quite similar to the volume obtained by accounting for the lattice collapse at each phase transition. The I-II, II-III, and III-V transitions produce volume collapses of 5.2, 3.6, and 2.3\% respectively, for a total of 10.7\% volume decrease as a result of structural change alone~\cite{HusbandBidDAC}. Applied to the Bi-I ambient pressure volume of 35.46~\AA{}$^3$, that gives us a value of 31.7~\AA$^3$ for Bi-V. Table~\ref{tab:BiEOS} shows the $K_0$ and $K_0'$ obtained using fits with each volume, a fit to the calculated data, as well as values from other works. The theoretical values in Table~\ref{tab:BiEOS} were obtained by a Vinet EOS fit to the calculated data points.

\begin{table}
	\centering
    \caption{Vinet equation of state parameters extracted from fits of all Bi data sets and our theoretical results. For fits stemming from this work, the data points from all three runs above 10~GPa were used. Uncertainty values are those obtained from the fit. Note that $V_0$ was fixed to the value obtained via theoretical calculations value for one of the fits, marked by an asterisk. Note that Ref.~\cite{LiuBiEOS} provides several sets of EOS parameters from different experiments; we list those done in Ar, which extended to highest pressure.}
    \label{tab:BiEOS}
    
\renewcommand{\arraystretch}{1}
\begin{tabular}{ | c | c | c | c | c |}
	\hline
	$K_0$ (GPa) & $K_0'$ & $V_0$ (\AA$^3$) & $P_{max}$ (GPa) & Ref. \\
	\hline
	38.2 $\pm$ 1.5 & 5.8 $\pm$ 0.06 & 31.67 $\pm$ 0.19 & 259 & Exp. \\
	\hline
	34.0 $\pm$ 0.2 & 6.0 $\pm$ 0.02 & 32.23$^{\ast}$ & 259 & Exp. \\
	\hline
	36.8 $\pm$ 1.0 & 6.0 $\pm$ 0.03 & 32.23 $\pm$ 0.13 & 462 & Th. \\
	\hline
	35.22 & 6.303 & 31.60 & 222 & \cite{AkahamaBiEOS} \\
	\hline
	42.7 & 5.3 & \textemdash{} & 52 & \cite{LiuBiEOS} \\
	\hline
	
\end{tabular}

\end{table}

\begin{figure}
    \centering
    \includegraphics[width=0.48\textwidth]{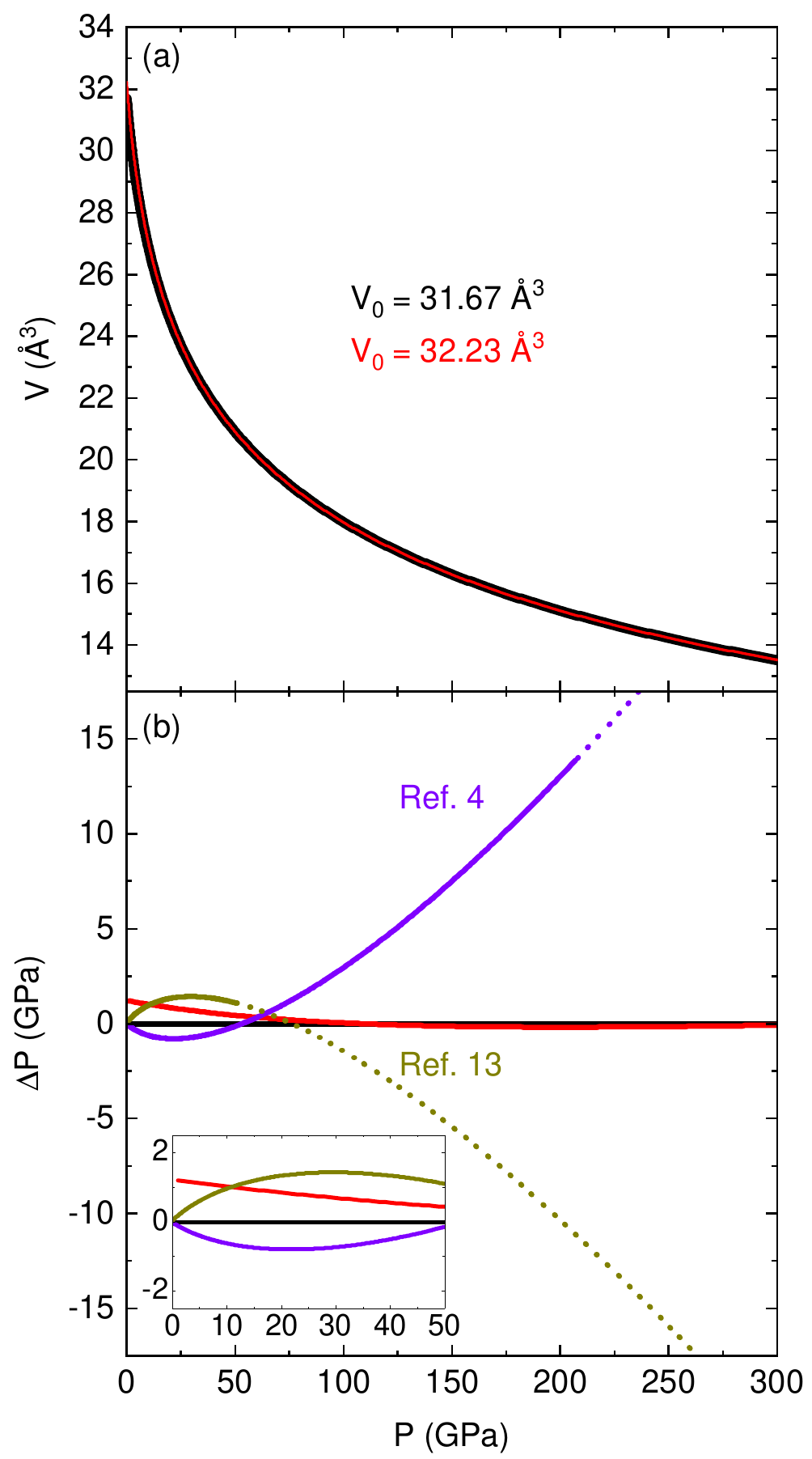}
    \caption{(a) The $P-V$ relation for Bi-V from Vinet equation of state fits to the data in Fig.~\ref{fig:PVAll} using two different ambient pressure volumes. Lines are different thicknesses as the data nearly completely overlap above 10~GPa. (b) The deviation from the fit with $V_0 = 31.67$~\AA$^3$, for the other choice of $V_0$ and literature values. The latter transition from solid to dotted lines above the maximum pressure reached in experiment. The inset shows a closeup of the 0-50~GPa range.}
    \label{fig:FitComp}
\end{figure}

In comparing two fits to the same experimental data, we find that while the choice of $V_0$ significantly changes $K_0$ and $K_0'$, it has less impact on the actual pressure-volume curves. In Fig.~\ref{fig:PVAll}(a) we use the values for $V_0 = 31.67$, but Fig.~\ref{fig:FitComp}(a) shows that the two curves are nearly indistinguishable over much of the range, as the changes in bulk modulus and derivative are compensated by the change in starting volume. The difference between the two only exceeds 1~GPa below 10~GPa, a pressure range where the Bi-V phase cannot be isolated [Fig.~\ref{fig:FitComp}(b)]. Meanwhile, comparison to previous work shows much more substantial spread, with a difference of more than 15~GPa from the EOS of Ref.~\cite{AkahamaBiEOS} by the maximum pressure reached in that study (220~GPa).

\section{Theoretical Results}

Density functional theory calculations for the atomic volume of Bi-V at various pressures have already been shown in Fig.~\ref{fig:PVAll}. In addition to that, we performed calculations of the energy difference between the bcc Bi-V structure and several other candidates: rhombohedral (Bi-I), monoclinic (Bi-II), fcc, sc, and hcp with either an ideal or relaxed $c/a$ axial ratio. Energies for the Bi-III structure were not calculated because of the complicated nature of the Bi-III host-guest structure and the fact that we mainly focus here on higher pressures. We choose to present the results as enthalpy ($H = E + PV$) differences of the various structures relative to the bcc (Bi-V) phase as functions of pressure.

\begin{figure}
    \centering
    \includegraphics[width=0.48\textwidth]{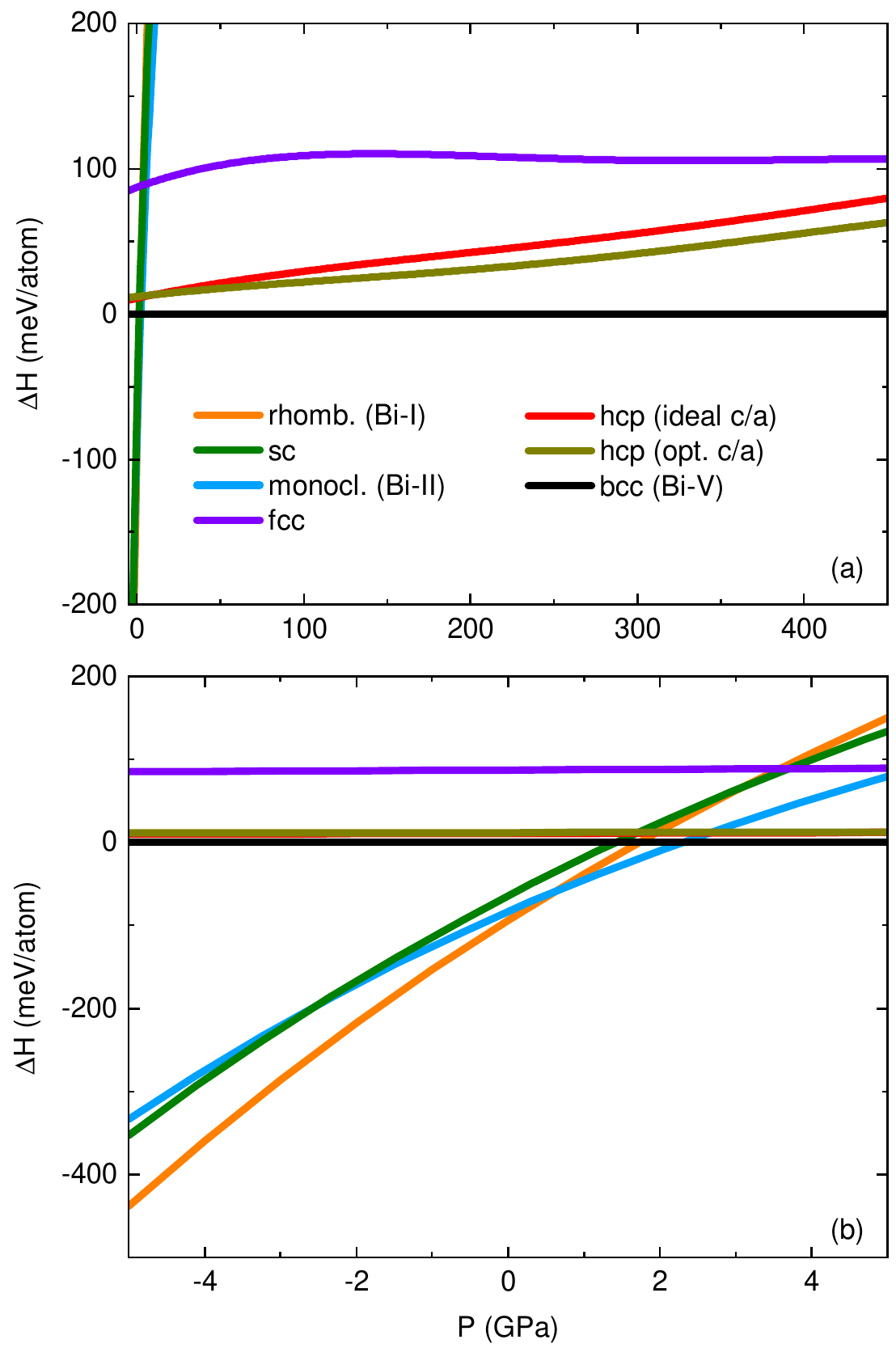}
    \caption{Calculated enthalpy differences of various potential structures of bismuth relative to bcc Bi-V (the solid line at $\Delta{}H = 0$) as functions of pressure. Panel (a) shows the entire pressure range over which calculations were performed. The bcc structure is clearly favored after only a slight increaes in pressure. Panel (b) inset shows a closeup of the low pressure region. the first pressure-induced phase (Bi-II) becomes favored over the ambient ground state (Bi-I) phase below 1~GPa. The Bi-III phase is not calculated (see Methods).}
    \label{fig:EDiffCalc}
\end{figure}

While our focus in this work is on the pressure range where the bcc phase is clearly favorable, we will comment briefly on the low pressure region where there is more competition [Fig.~\ref{fig:EDiffCalc}, inset]. The DFT model reproduces the correct ambient-pressure Bi-I phase and the pressure-induced transition to Bi-II. The reported transition takes place at around 2.5~GPa \cite{HusbandBidDAC} while our calculations indicate a transition below 1~GPa. The relatively small discrepancy is attributed to the fact that the phases are not fully optimized in the calculations. The transition to the bcc phase is predicted to occur too soon (about 2.3~GPa) but that is because we are not relaxing the Bi-I structure and also not including the Bi-III phase in our consideration. At multimegabar pressures bcc is comfortably stable over the two hcp phases, the next closest in energy, and the difference increases with pressure. Interestingly, the pressure behavior of fcc is rather similar to bcc but at an appreciably higher energy. At some high pressure, yet not studied, it is possible bismuth can transform to either hcp or fcc or perhaps a new phase altogether. 

\section{Evaluation of Nonhydrostatic Stress}

\begin{figure}
    \centering
    \includegraphics[width=0.48\textwidth]{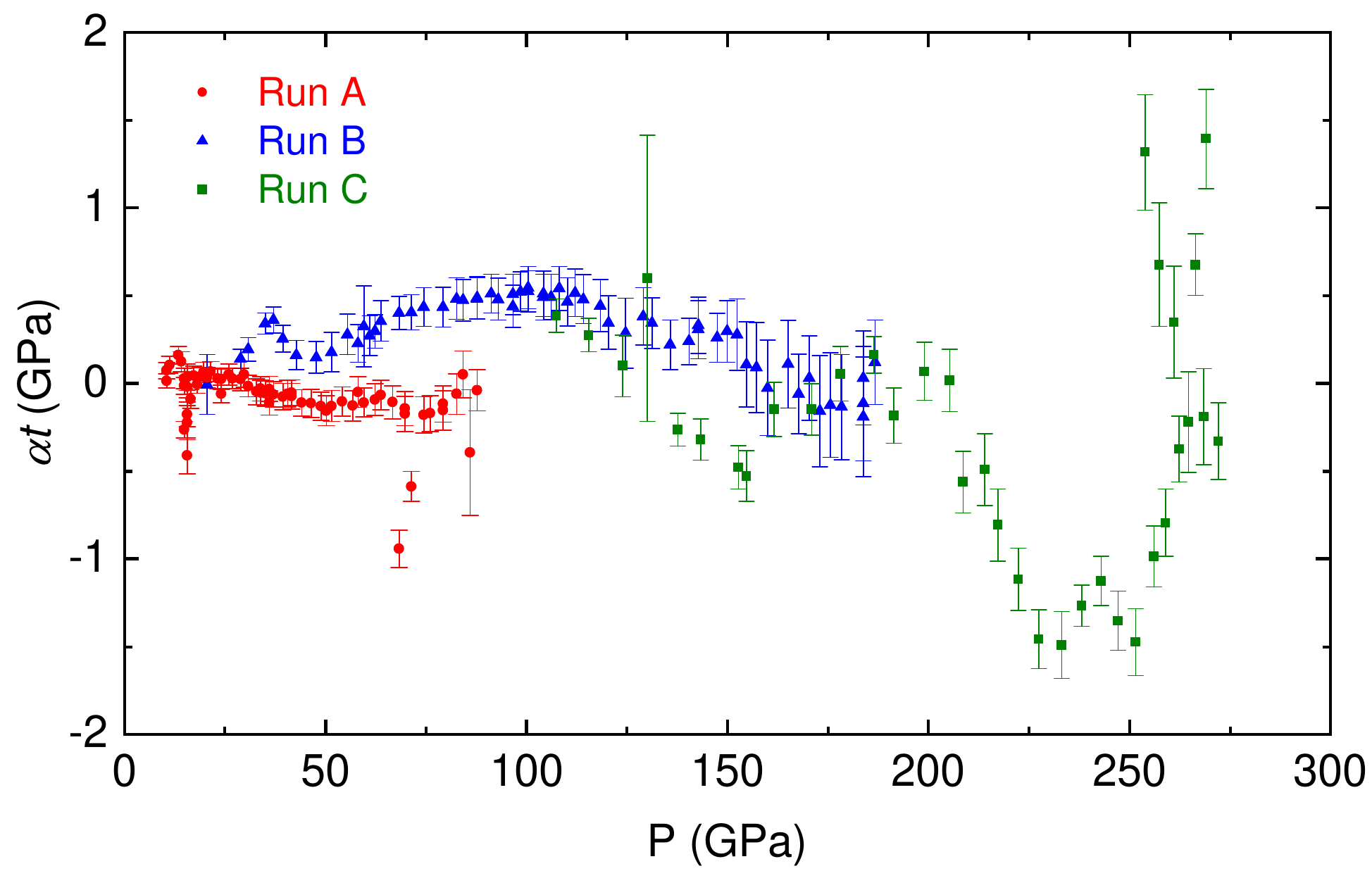}
    \caption{Values of $\alpha{}t$ of Bi, assumed to be the material strength, for all three experiments, using lineshift analysis as described in the text. Error values are derived from the uncertainty of the $d$-spacing from the peak fitting program.}
    \label{fig:Strength}
\end{figure}

It is possible to quantify the amount of uniaxial stress on the sample via the shift in \textit{d}-spacing for different \textit{HKL} values. Extensive discussion and analysis of this lineshift has been performed in other works~\cite{SinghGeneralNonhydrostat,AkahamaBiEOS,SinghNbNonhydrostat,SinghBiNonhydro}. For a cubic system, the basic equation for the lattice parameter $a_m$ is 

\begin{equation}
    a_m(h, k, l) = M_0 + M_1[3\Gamma{}(1-3\textrm{sin}^2\theta{})]
    \label{LPLineshift}
\end{equation}

with 

\begin{equation}
    \begin{aligned}
    M_0 = & a_P\{[1 + (\alpha{}t/3)(1 - 3\textrm{sin}^2\theta{}) \\
    & \times (S_{11} - S_{12}) - (1-\alpha{}^{-1})(2G_V)^{-1}\big]\}
    \end{aligned}
    \label{M0Lineshift}
\end{equation}

\begin{equation}
    M_1 = -a_P(\alpha{}St/3)
    \label{M1Lineshift}
\end{equation}

\begin{equation}
  \Gamma{}(h, k, l) = (h^2k^2 + k^2l^2 + h^2l^2)/(h^2 + k^2 + l^2)^2
    \label{GammaLineshift}
\end{equation}

where $a_P$ is the hydrostatic lattice parameter, $S_{ij}$ are the components of the compliance tensor, $S = S_{11} - S_{12} - S_{44}/2$, and $G_V$ is the shear modulus of randomly oriented polycrystals under isostrain conditions. $t$ is the quantity of interest, and represents $\sigma{}_3 - \sigma{}_1$, the difference between the radial and axial stress components. To simplify things, we can approximate $M_0 \simeq a_P$, meaning that at each pressure $a_m$ will be a linear function of $3(1 - 3\textrm{sin}^2\theta{})\Gamma{}$, with both $\theta{}$ and $\Gamma{}$ functions only of $h, k$ and $l$. Therefore, a linear fit of $a_m ~\textrm{vs.}~ 3(1 - 3\textrm{sin}^2\theta{})\Gamma{}$ gives us $M_0$ and $M_1$, and from Eq.~\ref{M1Lineshift} we can solve for $\alpha{}t = -3\frac{M_1}{M_0}\frac{1}{S}$. $\alpha{}$ determines the weighting of the shear moduli and takes on values ranging between 1 (for stress continuity) and 0.5 (halfway between stress and strain continuities). Since we cannot determine its value from our data, we present the combined product $\alpha{}t$ of our three data sets [Fig.~\ref{fig:Strength}]

Calculating $\alpha{}t$ requires values of the compliance tensor at each pressure point. Since these have not been measured experimentally up to such high pressures, we fit the theoretical results of Ref.~\cite{GutierrezBiCalcns} for the elastic tensor $C_{ij}$ up to 191~GPa to a power law of the form $C_{ij}(P) = C_{ij}(0) + aP^n$, with $n \simeq 0.9$ for each of $C_{11}, ~C_{12}, ~\textrm{and}~ C_{44}$. These values were then extrapolated up to 259~GPa and converted to the necessary compliance tensor results via $S_{ij} = C_{ij}^{-1}$.

Our three data sets are shown in Fig.~\ref{fig:Strength}, where the [110], [200], and [211] peak positions of Bi-V were used to construct gamma plots for each pressure and subsequently calculate $\alpha{}t$. Error bars were estimated from the uncertainty of the initial fitting of each peak. The maximum and minimum error in $a_{hkl}$ was calculated and used to bound the error of $M_1$ and subsequently $\alpha{}t$. Note that this does not account for other sources of uncertainty, for example the extrapolation of fits of the elastic constants used to calculate \textit{S}. The three data sets have three different trends. Run A has low values over its entire range, even becoming negative at higher pressure. For run B there is a peak in $\alpha{}t$ near 100~GPa. This may correspond to a deformation of the cell or gasket, as such a process is known to happen in the intermediate range of the loading curve, and the difference from run A could be the use of beveled anvils~\cite{LiDACBehavior}. Overall for both runs values are quite low, indicating good hydrostatic conditions. While error bars are large relative to the values of $\alpha{}t$, this is due to the similar values of $a_{hkl}$ for each peak, whose difference is only slightly larger than the uncertainty in peak positions.

Run C displays more variable behavior and higher absolute values, though like with the other data they are still less than 1\% of the total pressure [SM, Fig.~S2]. The larger spread in values can be linked to the reduced area covered by the detector (compare the two panels on the right hand side of Fig.~\ref{fig:XRD}) , making it harder to view complete rings on the detector that are necessary to precisely evaluate differential stress. It may also be due to challenges associated with the smaller culet size, and the inevitable differences in preparation of each experiment. A negative $\alpha{}t$ would correspond to the radial stress on the sample being larger than the axial stress, which is hard to reconcile with the fact that pressure is applied axially. While this is also the case with run A, for that experiment positive values are within the error and the overall conclusion is that $\alpha{}t$ is too small to be reliably measured. We note that $t$ is linearly related to the shear strength of the material. The high compressibility of Bi means it is unable to support a large deviatoric stress, keeping $t$ low. This is part of what makes it a good pressure calibrant. This was the argument used elsewhere for why there was no need for an even softer material when compressing Bi. However, our results for the first two runs are a little lower than those of other studies~\cite{LiuBiEOS,SinghBiNonhydro}, where $t$ values of around 0.5~GPa were seen by 200~GPa, though this is again hampered by the large relative uncertainty. We attribute  to the use of the soft Ne pressure medium, which is even more malleable than Bi and thus even better able to redistribute the applied stress, a conclusion backed up by the comparison of $P$($V$) data.

\section{Summary}

Our data show that by 260~GPa the volume of Bi is about 45\% of its estimated ambient pressure value. This can be compared to values for some common pressure markers: 70\%, 63\%, 58\%, and 40\% for Pt, Au, Cu, and NaCl (B2, CsCl-type phase), respectively~\cite{DorfmanIntercomparison,DewaeleEOSSixMetals}. Bismuth has a larger change in volume than the three coinage metals, which is especially beneficial to experimental precision at higher pressure, where $V(P$) flattens out. While not quite as compressible as NaCl, Bi has a much higher density and molar mass, leading to a stronger XRD signal. At high pressure, once it is fully in the bcc phase, Bi thus has advantages over the typically used calibrants.

In three different experiments, we have seen consistent behavior of the volume of the fcc Bi-V phase up to 260~GPa. These data show good overlap and can be readily fit to a Vinet equation of state. The upward deviation of previous \textit{P}(\textit{V}) data over 2~Mbar and the conclusions of lineshift analysis show that Ne better reduces anisotropic stress than Bi alone. As seen in Fig.~\ref{fig:FitComp}, at relatively low pressure the difference is minor. But when extending to the highest measured pressures or considering experiments beyond this range, the differences become significant. The information provided here can help motivate interest in Bi as a pressure calibrant above 10~GPa in XRD experiments. It also makes clear that the use of the softest pressure medium possible is necessary to reduce deviatoric stress and obtain accurate equations of state.

\section{Acknowledgments}

This work was performed under the auspices of the U.S. Department of Energy by Lawrence Livermore National Laboratory under Contract DE-AC52-07NA27344. Portions of this work were performed at HPCAT (Sector 16), Advanced Photon Source (APS), Argonne National Laboratory. HPCAT operations are supported by DOE-NNSA’s Office of Experimental Sciences. The Advanced Photon Source is a U.S. Department of Energy (DOE) Office of Science User Facility operated for the DOE Office of Science by Argonne National Laboratory under Contract No. DE-AC02-06CH11357.

\bibliography{Bi_EOSRefs}

\end{document}


\title{Supplemental material for: ``A new room-temperature equation of state of Bi up to 260 GPa''}

\author{Daniel J. Campbell}
\author{Daniel T. Sneed}
\author{E.F. O'Bannon III}
\author{Per S\"{o}derlind}
\author{Zsolt Jenei}
\affiliation{Lawrence Livermore National Laboratory, 7000 East Avenue, Livermore, CA 94550, USA}

\date{\today}

\maketitle

\section{Comparison of pressure scales}

As mentioned in the main text, we used the results of Run B, where Bi, Cu, and Ne were all present, to derive a Ne equation of state calibrated to Cu that could be used for Run C, where only Bi and Ne were present. In doing this we ensure that, despite using two different materials, our calibrations ultimately derive from the same source, namely Ref.~\cite{DewaeleEOSSixMetals}. By reducing the number of steps between the original calibration we reduce compounding error. Furthermore, our three experiments were all performed with similar setups and DACs. The resulting Ne EOS is in fact very similar to that of Ref.~\cite{DewaeleNeDiamondEOS}, which is the highest pressure Ne calibration available. Table~\ref{tab:BiDACRuns} compares our calibration with that one as well as another commonly used set of parameters from Fei et al.~\cite{FeiNeEOS}. However, the maximum pressure in the latter measurement was 115~GPa, which would require extending it to more than twice that value for our data sets, and its values are further from the other two. 

\begin{figure}
    \centering
    \includegraphics[width=0.5\textwidth]{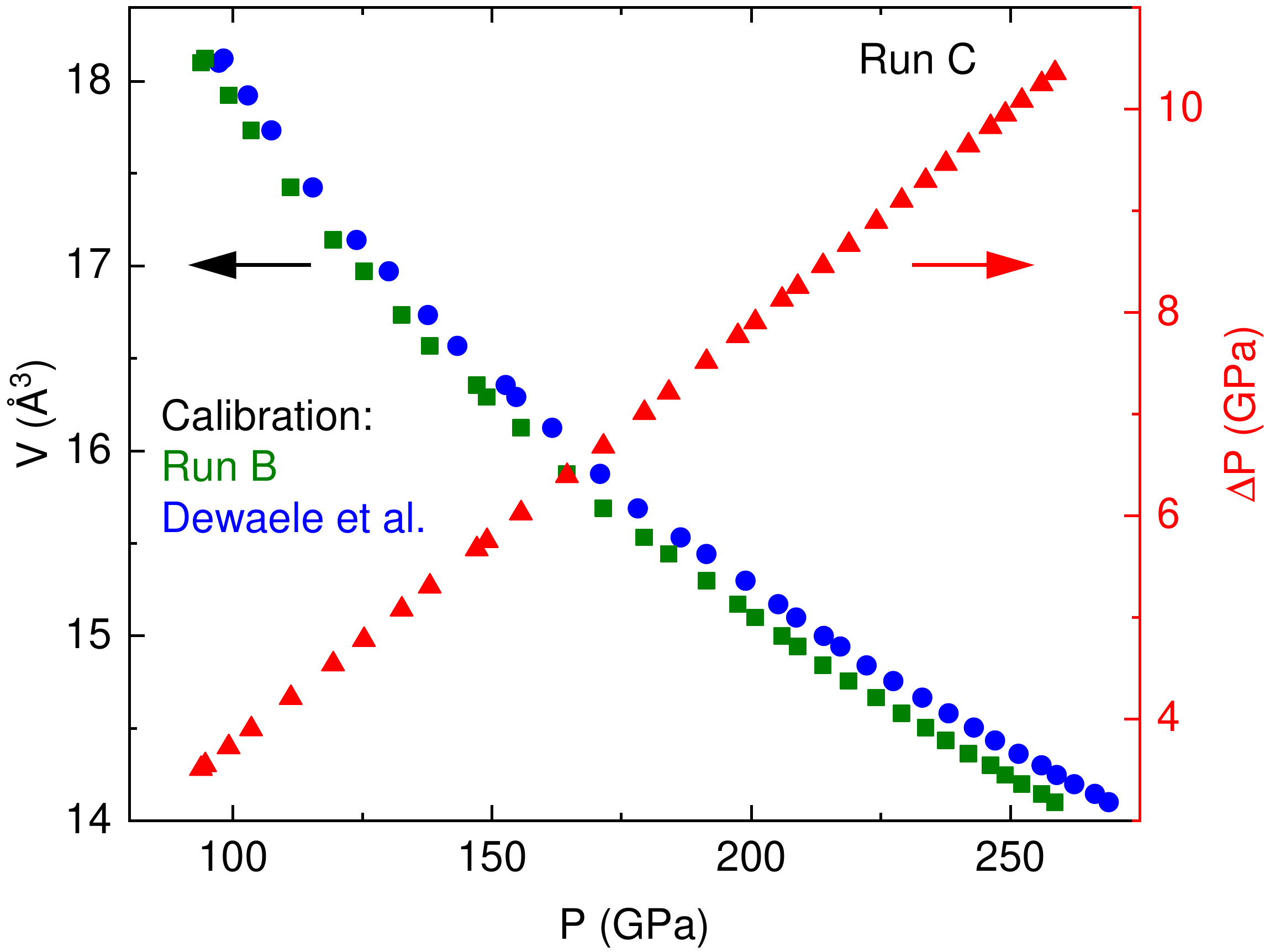}
    \caption{A comparison of two different pressure scales for the volume of Bi in the Run C data. On the left axis, the calibration determined by comparing to Cu in the Run B data (green squares) and that of Dewaele et al.~\cite{DewaeleNeDiamondEOS} (blue circles). The right y-axis shows the difference in pressure between the two calibrations (red triangles)}
    \label{fig:EOSComp}
\end{figure}

Despite the similarities in the values from our data and those of Ref.~\cite{DewaeleNeDiamondEOS}, by the highest pressure of Run C they still differ by 10~GPa, or about 4\% [Fig.~\ref{fig:EOSComp}]. Since the same $V_0$ was used and the two sets of $K_0$ and $K_0'$ values are very similar, the difference between the two Vinet equations grows essentially linearly with pressure, underscoring the importance of consistency between calibrations for experiments far beyond 1~Mbar. We note that the calibration through our own data results in a lower volume for a given pressure. This is further support for using our calibration, as the lower volume indicates reduced uniaxial stress.

\begin{table}[h]
	\centering
    \caption{Three different equation of state parameters for Ne. The 0~GPa atomic volume was fixed to 22.234~\AA$^3$ for the Run B data to match the value of Ref.~\cite{DewaeleNeDiamondEOS} (In Ref.~\cite{FeiNeEOS} it is 22.241~\AA$^3$).\newline}
    \label{tab:BiDACRuns}
    
\begin{tabular}{ | c | c | c | c | c |}
	\hline
	$K_0$ (GPa) & $K_0'$ & $P_{max}$ (GPa) & Ref. \\
	\hline
	1.046 & 8.38 & 184 & Run B \\
	\hline
	1.070 & 8.40 & 209 & \cite{DewaeleNeDiamondEOS} \\
	\hline
    1.16 & 8.23 & 115 & \cite{FeiNeEOS} \\
	\hline
	
\end{tabular}

\end{table}

\newpage

\section{Relative Uniaxial Stress Component}

\begin{figure}[h]
    \centering
    \includegraphics[width=0.6\textwidth]{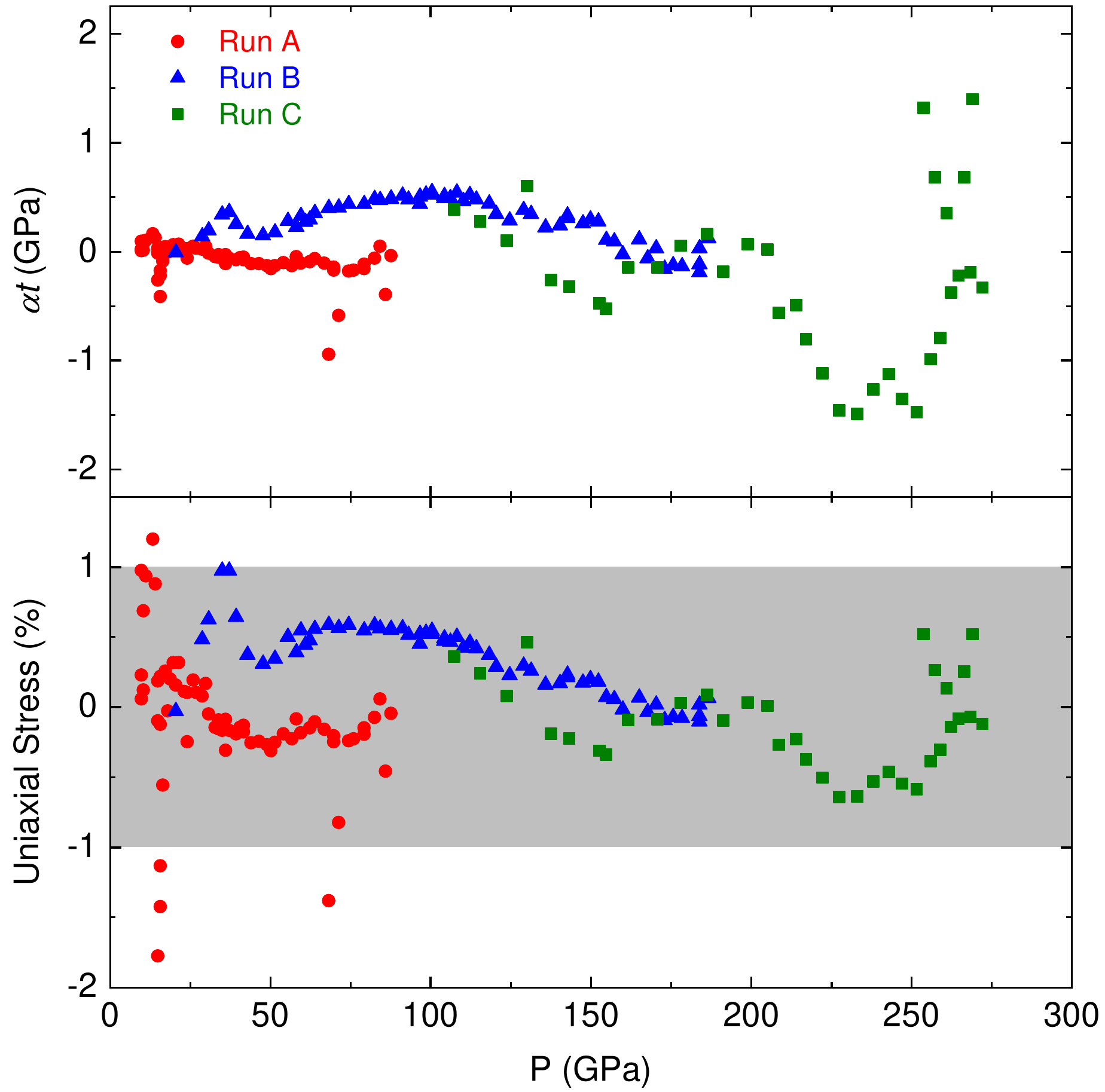}
    \caption{(a) The absolute value of $\alpha{}t$ for the three experiments, the same data presented in the main text. (b) The same values as a percentage of the total pressure measured by the Cu or Ne lattice parameter. The region within $\pm$1\% is highlighted in gray.}
    \label{fig:RelativeStrength}
\end{figure}

Figure~\ref{fig:RelativeStrength}(a) reproduces Fig.~5 from the main text, with error bars removed for clarity. The lower panel shows those same data relative to the total pressure. It can be seen that only for a few data points do the values exceed $\pm$1\% of the total pressure as measured by either Cu or Ne.

\clearpage

\section{Pressure-Volume data for Bi-V}

\begin{table}[h]
\centering
\caption{Lattice parameters (in \AA{}) of Bi and (when present) Cu and Ne during the three experiments, as well as calculated values for Bi. As in the main text, pressures (in GPa) are calculated with the Cu EOS~\cite{DewaeleEOSSixMetals} for Runs A and B, and our Ne EOS for Run C. The missing Ne value in one case for Run B is because a Ne peak could not be found in that pattern. Data are presented only above 10~GPa, when no peak from the Bi-III phase are observed.\newline}
\begin{tabular}{|c|c|c|c|c|} 
\hline
Run & P & $a_{\textrm{Cu}}$ & $a_{\textrm{Ne}}$               & $a_{\textrm{Bi-V}}$  \\ 
\hline
A   & 10.07                        & 3.5386                             & --                          & 3.7732                              \\ 
\hline
A   & 10.15                        & 3.5382                             & --                          & 3.7709                              \\ 
\hline
A   & 10.15                        & 3.5382                             & --                          & 3.7714                              \\ 
\hline
A   & 10.37                        & 3.5368                             & --                          & 3.7688                              \\ 
\hline
A   & 10.69                        & 3.5347                             & --                          & 3.7642                              \\ 
\hline
A   & 11.54                        & 3.5294                             & --                          & 3.7527                              \\ 
\hline
A   & 13.61                        & 3.5170                             & --                          & 3.7278                              \\ 
\hline
A   & 14.18                        & 3.5137                             & --                          & 3.7217                              \\ 
\hline
A   & 14.73                        & 3.5105                             & --                          & 3.7177                              \\ 
\hline
A   & 15.10                        & 3.5084                             & --                          & 3.7142                              \\ 
\hline
A   & 15.38                        & 3.5068                             & --                          & 3.7112                              \\ 
\hline
A   & 15.86                        & 3.5042                             & --                          & 3.7052                              \\ 
\hline
A   & 15.88                        & 3.5041                             & --                          & 3.7037                              \\ 
\hline
A   & 15.94                        & 3.5037                             & --                          & 3.7058                              \\ 
\hline
A   & 15.95                        & 3.5037                             & --                          & 3.7052                              \\ 
\hline
A   & 16.10                        & 3.5028                             & --                          & 3.7022                              \\ 
\hline
A   & 16.37                        & 3.5014                             & --                          & 3.6999                              \\ 
\hline
A   & 16.85                        & 3.4987                             & --                          & 3.6943                              \\ 
\hline
A   & 17.64                        & 3.4945                             & --                          & 3.6839                              \\ 
\hline
A   & 18.47                        & 3.4901                             & --                          & 3.6759                              \\ 
\hline
A   & 19.29                        & 3.4859                             & --                          & 3.6668                              \\ 
\hline
A   & 20.16                        & 3.4814                             & --                          & 3.6581                              \\ 
\hline
A   & 21.21                        & 3.4763                             & --                          & 3.6484                              \\ 
\hline
A   & 22.30                        & 3.4709                             & --                          & 3.6380                              \\ 
\hline
A   & 23.58                        & 3.4648                             & --                          & 3.6276                              \\ 
\hline
A   & 24.98                        & 3.4583                             & --                          & 3.6166                              \\ 
\hline
A   & 26.30                        & 3.4524                             & --                          & 3.6062                              \\ 
\hline
A   & 27.71                        & 3.4462                             & --                          & 3.5958                              \\ 
\hline
A   & 29.54                        & 3.4383                             & --                          & 3.5827                              \\ 
\hline
A   & 30.70                        & 3.4335                             & --                          & 3.5751                              \\ 
\hline
A   & 31.89                        & 3.4286                             & --                          & 3.5678                              \\ 
\hline
A   & 33.26                        & 3.4231                             & --                          & 3.5606                              \\ 
\hline
A   & 34.46                        & 3.4184                             & --                          & 3.5550                              \\ 
\hline
A   & 34.67                        & 3.4176                             & --                          & 3.5541                              \\ 
\hline
A   & 35.17                        & 3.4156                             & --                          & 3.5514                              \\ 
\hline
A   & 35.92                        & 3.4128                             & --                          & 3.5468                              \\ 
\hline
A   & 36.99                        & 3.4087                             & --                          & 3.5403                              \\ 
\hline
A   & 38.18                        & 3.4043                             & --                          & 3.5328                              \\ 
\hline
A   & 39.47                        & 3.3996                             & --                          & 3.5251                              \\ 
\hline
A   & 40.76                        & 3.3950                             & --                          & 3.5173                              \\ 
\hline
A   & 41.88                        & 3.3911                             & --                          & 3.5110                              \\ 
\hline
A   & 42.78                        & 3.3879                             & --                          & 3.5062                              \\ 
\hline
A   & 44.35                        & 3.3826                             & --                          & 3.4978                              \\ 
\hline
A   & 47.02                        & 3.3737                             & --                          & 3.4847                              \\ 
\hline
A   & 49.27                        & 3.3665                             & --                          & 3.4737                              \\ 
\hline
A   & 51.30                        & 3.3601                             & --                          & 3.4639                              \\ 
\hline

\end{tabular}
\end{table}
\begin{table}
\centering
\begin{tabular}{|c|c|c|c|c|} 

\hline
Run & P & $a_{\textrm{Cu}}$ & $a_{\textrm{Ne}}$               & $a_{\textrm{Bi-V}}$  \\ 
\hline
A   & 52.70                        & 3.3558                             & --                          & 3.4574                              \\ 
\hline
A   & 54.38                        & 3.3508                             & --                          & 3.4503                              \\ 
\hline
A   & 56.67                        & 3.3440                             & --                          & 3.4409                              \\ 
\hline
A   & 58.72                        & 3.3381                             & --                          & 3.4319                              \\ 
\hline
A   & 60.80                        & 3.3323                             & --                          & 3.4242                              \\ 
\hline
A   & 62.72                        & 3.3270                             & --                          & 3.4165                              \\ 
\hline
A   & 64.92                        & 3.3211                             & --                          & 3.4084                              \\ 
\hline
A   & 66.86                        & 3.3160                             & --                          & 3.4014                              \\ 
\hline
A   & 69.57                        & 3.3090                             & --                          & 3.3951                              \\ 
\hline
A   & 69.62                        & 3.3089                             & --                          & 3.3963                              \\ 
\hline
A   & 70.87                        & 3.3057                             & --                          & 3.3923                              \\ 
\hline
A   & 72.98                        & 3.3005                             & --                          & 3.3909                              \\ 
\hline
A   & 74.84                        & 3.2959                             & --                          & 3.3793                              \\ 
\hline
A   & 77.45                        & 3.2897                             & --                          & 3.3712                              \\ 
\hline
A   & 80.22                        & 3.2832                             & --                          & 3.3627                              \\ 
\hline
A   & 80.57                        & 3.2824                             & --                          & 3.3615                              \\ 
\hline
A   & 82.84                        & 3.2773                             & --                          & 3.3544                              \\ 
\hline
A   & 85.63                        & 3.2711                             & --                          & 3.3466                              \\ 
\hline
A   & 88.59                        & 3.2646                             & --                          & 3.3385                              \\ 
\hline
\hline
\hline
B   & 16.80                        & 3.4990                             & \multicolumn{1}{r|}{3.4333} & 3.6918                              \\ 
\hline
B   & 17.48                        & 3.4953                             & \multicolumn{1}{r|}{3.4240} & 3.6865                              \\ 
\hline
B   & 14.34                        & 3.5127                             & \multicolumn{1}{r|}{3.4154} & 3.6820                              \\ 
\hline
B   & 17.99                        & 3.4926                             & \multicolumn{1}{r|}{3.4144} & 3.6807                              \\ 
\hline
B   & 18.76                        & 3.4886                             & \multicolumn{1}{r|}{3.4079} & 3.6772                              \\ 
\hline
B   & 19.50                        & 3.4848                             & --                          & 3.6768                              \\ 
\hline
B   & 18.55                        & 3.4897                             & \multicolumn{1}{r|}{3.4062} & 3.6752                              \\ 
\hline
B   & 18.61                        & 3.4894                             & \multicolumn{1}{r|}{3.4036} & 3.6751                              \\ 
\hline
B   & 18.74                        & 3.4887                             & \multicolumn{1}{r|}{3.4024} & 3.6743                              \\ 
\hline
B   & 19.29                        & 3.4859                             & \multicolumn{1}{r|}{3.3949} & 3.6675                              \\ 
\hline
B   & 21.73                        & 3.4737                             & \multicolumn{1}{r|}{3.3616} & 3.6452                              \\ 
\hline
B   & 28.94                        & 3.4408                             & \multicolumn{1}{r|}{3.2814} & 3.5880                              \\ 
\hline
B   & 31.31                        & 3.4310                             & \multicolumn{1}{r|}{3.2608} & 3.5728                              \\ 
\hline
B   & 36.10                        & 3.4121                             & \multicolumn{1}{r|}{3.2230} & 3.5433                              \\ 
\hline
B   & 38.01                        & 3.4049                             & \multicolumn{1}{r|}{3.2082} & 3.5273                              \\ 
\hline
B   & 39.83                        & 3.3983                             & \multicolumn{1}{r|}{3.1946} & 3.5162                              \\ 
\hline
B   & 42.93                        & 3.3874                             & \multicolumn{1}{r|}{3.1737} & 3.4993                              \\ 
\hline
B   & 47.37                        & 3.3726                             & \multicolumn{1}{r|}{3.1463} & 3.4769                              \\ 
\hline
B   & 51.71                        & 3.3588                             & \multicolumn{1}{r|}{3.1229} & 3.4568                              \\ 
\hline
B   & 54.81                        & 3.3495                             & \multicolumn{1}{r|}{3.1072} & 3.4432                              \\ 
\hline
B   & 56.01                        & 3.3459                             & \multicolumn{1}{r|}{3.0990} & 3.4371                              \\ 
\hline
B   & 58.88                        & 3.3377                             & \multicolumn{1}{r|}{3.0879} & 3.4269                              \\ 
\hline
B   & 59.98                        & 3.3346                             & \multicolumn{1}{r|}{3.0830} & 3.4232                              \\ 
\hline
B   & 60.92                        & 3.3320                             & \multicolumn{1}{r|}{3.0785} & 3.4192                              \\ 
\hline
B   & 62.62                        & 3.3273                             & \multicolumn{1}{r|}{3.0728} & 3.4134                              \\ 
\hline
B   & 67.68                        & 3.3138                             & \multicolumn{1}{r|}{3.0554} & 3.3973                              \\ 
\hline
B   & 71.07                        & 3.3052                             & \multicolumn{1}{r|}{3.0432} & 3.3849                              \\ 
\hline
B   & 74.35                        & 3.2971                             & \multicolumn{1}{r|}{3.0321} & 3.3744                              \\ 
\hline
B   & 79.14                        & 3.2858                             & \multicolumn{1}{r|}{3.0182} & 3.3603                              \\ 
\hline
B   & 81.81                        & 3.2796                             & \multicolumn{1}{r|}{3.0096} & 3.3516                              \\ 
\hline
B   & 83.91                        & 3.2749                             & \multicolumn{1}{r|}{3.0027} & 3.3450                              \\ 
\hline
B   & 85.79                        & 3.2707                             & \multicolumn{1}{r|}{2.9981} & 3.3405                              \\ 
\hline
B   & 86.62                        & 3.2689                             & \multicolumn{1}{r|}{2.9943} & 3.3370                              \\ 
\hline
B   & 89.34                        & 3.2630                             & \multicolumn{1}{r|}{2.9874} & 3.3302                              \\ 
\hline
B   & 91.46                        & 3.2585                             & \multicolumn{1}{r|}{2.9813} & 3.3244                              \\ 
\hline
B   & 94.55                        & 3.2521                             & \multicolumn{1}{r|}{2.9737} & 3.3164                              \\ 
\hline
B   & 96.14                        & 3.2489                             & \multicolumn{1}{r|}{2.9729} & 3.3135                              \\ 
\hline
B   & 97.79                        & 3.2455                             & \multicolumn{1}{r|}{2.9673} & 3.3084                              \\ 
\hline
B   & 100.14                       & 3.2409                             & \multicolumn{1}{r|}{2.9623} & 3.3037                              \\ 
\hline
B   & 101.25                       & 3.2387                             & \multicolumn{1}{r|}{2.9591} & 3.3006                              \\ 
\hline
B   & 104.04                       & 3.2333                             & \multicolumn{1}{r|}{2.9531} & 3.2946                              \\ 
\hline
B   & 105.08                       & 3.2313                             & \multicolumn{1}{r|}{2.9501} & 3.2915                              \\ 
\hline

\end{tabular}
\end{table}
\begin{table}
\centering
\begin{tabular}{|c|c|c|c|c|} 
\hline
Run & P & $a_{\textrm{Cu}}$ & $a_{\textrm{Ne}}$               & $a_{\textrm{Bi-V}}$  \\ 
\hline
B   & 107.16                       & 3.2274                             & \multicolumn{1}{r|}{2.9450} & 3.2864                              \\ 
\hline
B   & 109.10                       & 3.2238                             & \multicolumn{1}{r|}{2.9405} & 3.2820                              \\ 
\hline
B   & 110.05                       & 3.2221                             & \multicolumn{1}{r|}{2.9375} & 3.2796                              \\ 
\hline
B   & 111.87                       & 3.2187                             & \multicolumn{1}{r|}{2.9337} & 3.2754                              \\ 
\hline
B   & 114.62                       & 3.2138                             & \multicolumn{1}{r|}{2.9277} & 3.2691                              \\ 
\hline
B   & 118.05                       & 3.2077                             & \multicolumn{1}{r|}{2.9194} & 3.2614                              \\ 
\hline
B   & 120.58                       & 3.2034                             & \multicolumn{1}{r|}{2.9138} & 3.2558                              \\ 
\hline
B   & 123.37                       & 3.1986                             & \multicolumn{1}{r|}{2.9077} & 3.2494                              \\ 
\hline
B   & 127.59                       & 3.1916                             & \multicolumn{1}{r|}{2.8994} & 3.2410                              \\ 
\hline
B   & 131.25                       & 3.1857                             & \multicolumn{1}{r|}{2.8918} & 3.2333                              \\ 
\hline
B   & 134.88                       & 3.1799                             & \multicolumn{1}{r|}{2.8843} & 3.2256                              \\ 
\hline
B   & 139.01                       & 3.1735                             & \multicolumn{1}{r|}{2.8762} & 3.2176                              \\ 
\hline
B   & 140.70                       & 3.1709                             & \multicolumn{1}{r|}{2.8733} & 3.2140                              \\ 
\hline
B   & 142.13                       & 3.1687                             & \multicolumn{1}{r|}{2.8695} & 3.2099                              \\ 
\hline
B   & 144.91                       & 3.1645                             & \multicolumn{1}{r|}{2.8634} & 3.2048                              \\ 
\hline
B   & 148.26                       & 3.1596                             & \multicolumn{1}{r|}{2.8575} & 3.1986                              \\ 
\hline
B   & 151.00                       & 3.1556                             & \multicolumn{1}{r|}{2.8508} & 3.1918                              \\ 
\hline
B   & 151.79                       & 3.1545                             & \multicolumn{1}{r|}{2.8483} & 3.1897                              \\ 
\hline
B   & 154.59                       & 3.1505                             & \multicolumn{1}{r|}{2.8442} & 3.1849                              \\ 
\hline
B   & 158.24                       & 3.1454                             & \multicolumn{1}{r|}{2.8377} & 3.1787                              \\ 
\hline
B   & 161.35                       & 3.1411                             & \multicolumn{1}{r|}{2.8328} & 3.1740                              \\ 
\hline
B   & 164.35                       & 3.1370                             & \multicolumn{1}{r|}{2.8272} & 3.1686                              \\ 
\hline
B   & 167.86                       & 3.1323                             & \multicolumn{1}{r|}{2.8214} & 3.1627                              \\ 
\hline
B   & 171.20                       & 3.1279                             & \multicolumn{1}{r|}{2.8167} & 3.1574                              \\ 
\hline
B   & 174.22                       & 3.1240                             & \multicolumn{1}{r|}{2.8121} & 3.1526                              \\ 
\hline
B   & 176.92                       & 3.1206                             & \multicolumn{1}{r|}{2.8076} & 3.1484                              \\ 
\hline
B   & 178.40                       & 3.1187                             & \multicolumn{1}{r|}{2.8046} & 3.1453                              \\ 
\hline
B   & 179.91                       & 3.1168                             & \multicolumn{1}{r|}{2.8032} & 3.1434                              \\ 
\hline
B   & 182.22                       & 3.1139                             & \multicolumn{1}{r|}{2.7993} & 3.1399                              \\ 
\hline
B   & 184.47                       & 3.1111                             & \multicolumn{1}{r|}{2.7955} & 3.1362                              \\ 
\hline
\hline
\hline
C   & 93.85                        & --             & \multicolumn{1}{r|}{2.9762} & 3.3080                              \\ 
\hline
C   & 94.69                        & --             & \multicolumn{1}{r|}{2.9739} & 3.3093                              \\ 
\hline
C   & 99.18                        & --             & \multicolumn{1}{r|}{2.9618} & 3.2972                              \\ 
\hline
C   & 103.54                       & --             & \multicolumn{1}{r|}{2.9505} & 3.2856                              \\ 
\hline
C   & 111.21                       & --             & \multicolumn{1}{r|}{2.9318} & 3.2664                              \\ 
\hline
C   & 119.39                       & --             & \multicolumn{1}{r|}{2.9132} & 3.2484                              \\ 
\hline
C   & 125.29                       & --             & \multicolumn{1}{r|}{2.9005} & 3.2378                              \\ 
\hline
C   & 132.56                       & --             & \multicolumn{1}{r|}{2.8857} & 3.2226                              \\ 
\hline
C   & 138.05                       & --             & \multicolumn{1}{r|}{2.8750} & 3.2120                              \\ 
\hline
C   & 147.03                       & --             & \multicolumn{1}{r|}{2.8585} & 3.1981                              \\ 
\hline
C   & 149.00                       & --             & \multicolumn{1}{r|}{2.8549} & 3.1940                              \\ 
\hline
C   & 155.60                       & --             & \multicolumn{1}{r|}{2.8435} & 3.1830                              \\ 
\hline
C   & 164.50                       & --             & \multicolumn{1}{r|}{2.8289} & 3.1665                              \\ 
\hline
C   & 171.46                       & --             & \multicolumn{1}{r|}{2.8180} & 3.1540                              \\ 
\hline
C   & 179.40                       & --             & \multicolumn{1}{r|}{2.8061} & 3.1435                              \\ 
\hline
C   & 184.15                       & --             & \multicolumn{1}{r|}{2.7992} & 3.1374                              \\ 
\hline
C   & 191.41                       & --             & \multicolumn{1}{r|}{2.7890} & 3.1276                              \\ 
\hline
C   & 197.48                       & --             & \multicolumn{1}{r|}{2.7807} & 3.1189                              \\ 
\hline
C   & 200.78                       & --             & \multicolumn{1}{r|}{2.7764} & 3.1139                              \\ 
\hline
C   & 205.96                       & --            & \multicolumn{1}{r|}{2.7697} & 3.1071                              \\ 
\hline
C   & 208.98                       & --             & \multicolumn{1}{r|}{2.7658} & 3.1032                              \\ 
\hline
C   & 213.89                       & --             & \multicolumn{1}{r|}{2.7597} & 3.0961                              \\ 
\hline
C   & 218.83                       & --             & \multicolumn{1}{r|}{2.7537} & 3.0903                              \\ 
\hline
C   & 224.22                       & --             & \multicolumn{1}{r|}{2.7472} & 3.0839                              \\ 
\hline
C   & 229.06                       & --             & \multicolumn{1}{r|}{2.7416} & 3.0779                              \\ 
\hline
C   & 233.69                       & --             & \multicolumn{1}{r|}{2.7363} & 3.0725                              \\ 
\hline
C   & 237.62                       & --             & \multicolumn{1}{r|}{2.7319} & 3.0677                              \\ 
\hline
C   & 241.96                       & --             & \multicolumn{1}{r|}{2.7271} & 3.0625                              \\ 
\hline

\end{tabular}
\end{table}

\begin{table}
\centering
\begin{tabular}{|c|c|c|c|c|} 
\hline
Run & P & $a_{\textrm{Cu}}$ & $a_{\textrm{Ne}}$               & $a_{\textrm{Bi-V}}$  \\ 
\hline
C   & 246.25                       & --           & \multicolumn{1}{r|}{2.7225} & 3.0581                              \\ 
\hline
C   & 249.09                       & --            & \multicolumn{1}{r|}{2.7195} & 3.0543                              \\ 
\hline
C   & 252.30                       & --             & \multicolumn{1}{r|}{2.7161} & 3.0507                              \\ 
\hline
C   & 256.12                       & --             & \multicolumn{1}{r|}{2.7121} & 3.0471                              \\ 
\hline
C   & 258.69                       & --             & \multicolumn{1}{r|}{2.7095} & 3.0439                              \\ 
\hline
\hline
\hline
Th. & -1.08                        & --             & --                          & 2.90186                             \\ 
\hline
Th. & -0.29                        & --             & --                          & 3.99808                             \\ 
\hline
Th. & 0.57                         & --             & --                          & 3.97657                             \\ 
\hline
Th. & 1.50                         & --             & --                          & 3.95508                             \\ 
\hline
Th. & 2.51                         & --             & --                          & 3.93357                             \\ 
\hline
Th. & 3.60                         & --             & --                          & 3.91208                             \\ 
\hline
Th. & 4.79                         & --             & --                          & 3.89058                             \\ 
\hline
Th. & 6.07                         & --             & --                          & 3.8691                              \\ 
\hline
Th. & 7.45                         & --             & --                          & 3.8476                              \\ 
\hline
Th. & 8.95                         & --             & --                          & 3.82609                             \\ 
\hline
Th. & 10.56                        & --             & --                          & 3.80461                             \\ 
\hline
Th. & 12.31                        & --             & --                          & 3.78312                             \\ 
\hline
Th. & 14.19                        & --             & --                          & 3.76162                             \\ 
\hline
Th. & 16.23                        & --             & --                          & 3.7401                              \\ 
\hline
Th. & 18.42                        & --             & --                          & 3.71863                             \\ 
\hline
Th. & 20.79                        & --             & --                          & 3.69715                             \\ 
\hline
Th. & 23.35                        & --             & --                          & 3.67566                             \\ 
\hline
Th. & 26.10                        & --             & --                          & 3.65417                             \\ 
\hline
Th. & 29.05                        & --             & --                          & 3.63268                             \\ 
\hline
Th. & 32.25                        & --             & --                          & 3.61118                             \\ 
\hline
Th. & 35.69                        & --             & --                          & 3.58968                             \\ 
\hline
Th. & 39.39                        & --             & --                          & 3.56819                             \\ 
\hline
Th. & 43.38                        & --             & --                          & 3.54669                             \\ 
\hline
Th. & 47.67                        & --             & --                          & 3.52515                             \\ 
\hline
Th. & 52.29                        & --             & --                          & 3.50367                             \\ 
\hline
Th. & 57.27                        & --             & --                          & 3.4822                              \\ 
\hline
Th. & 62.64                        & --             & --                          & 3.46068                             \\ 
\hline
Th. & 68.41                        & --             & --                          & 3.43922                             \\ 
\hline
Th. & 74.63                        & --            & --                          & 3.41773                             \\ 
\hline
Th. & 81.33                        & --             & --                          & 3.39619                             \\ 
\hline
Th. & 88.54                        & --            & --                          & 3.37473                             \\ 
\hline
Th. & 96.29                        & --             & --                          & 3.35322                             \\ 
\hline
Th. & 104.63                       & --            & --                          & 3.33174                             \\ 
\hline
Th. & 113.61                       & --            & --                          & 3.31022                             \\ 
\hline
Th. & 123.27                       &  --          & --                          & 3.28873                             \\ 
\hline
Th. & 133.65                       & --             & --                          & 3.26726                             \\ 
\hline
Th. & 144.83                       & --            & --                          & 3.24576                             \\ 
\hline
Th. & 156.86                       & --             & --                          & 3.22423                             \\ 
\hline
Th. & 169.81                       & --             & --                          & 3.20273                             \\ 
\hline
Th. & 183.73                       & --             & --                          & 3.18127                             \\ 
\hline
Th. & 198.72                       & --            & --                          & 3.15978                             \\ 
\hline
Th. & 214.86                       & --             & --                          & 3.13827                             \\ 
\hline
Th. & 232.26                       & --            & --                          & 3.1168                              \\ 
\hline
Th. & 250.97                       & --             & --                          & 3.09531                             \\ 
\hline
Th. & 271.09                       & --            & --                          & 3.0738                              \\ 
\hline
Th. & 292.73                       & --             & --                          & 3.05227                             \\ 
\hline
Th. & 316.50                       & --             & --                          & 3.03079                             \\ 
\hline
Th. & 342.01                       & --             & --                          & 3.0093                              \\ 
\hline
Th. & 368.83                       & --             & --                          & 2.9878                              \\ 
\hline
Th. & 397.68                       & --            & --                          & 2.96629                             \\ 
\hline
Th. & 428.75                       & --         & --                          & 2.94484                             \\ 
\hline
Th. & 462.36                       & --             & --                          & 2.92332                             \\
\hline
\end{tabular}
\end{table}

\newpage

\bibliography{Bi_EOSRefs}